\documentclass[twocolumn,twoside,slac_two]{revtex4}
\usepackage{graphicx}
\usepackage{fancyhdr}
\newcommand{\apj}{ApJ}
\newcommand{\apjl}{ApJL}
\newcommand{\apjs}{ApJS}

\newcommand{\apss}{Ap\&SS}
\newcommand{\aap}{A\&A}

\newcommand{\aaps}{A\&AS}

\newcommand{\mnras}{MNRAS}

\newcommand{\pasj}{PASJ}
\newcommand{\nat}{Nature}

\begin{document}

\title{ 
Study of luminosity and spin-up relation in X-ray binary pulsars with
long-term monitoring by MAXI/GSC and Fermi/GBM
}

%

\author{Mutsumi Sugizaki, Tatehiro Mihara}
\affiliation{
Institute of Physical and Chemical Research (RIKEN), 2-1 Hirosawa, Wako, Saitama 351-0198, Japan
}
%
\author{Motoki Nakajima}
\affiliation{
Nihon University, 2-870-1 Sakaecho-nishi, Matsudo, Chiba 101-830, Japan
}

\author{Kazutaka Yamaoka}
\affiliation{
Nagoya University, Furo-cho, Chikusa-ku, Nagoya 464-8601, Japan
}




\begin{abstract}

We study the relation between luminosity and spin-period change
in X-ray binary pulsars using long-term light curve obtained
by the MAXI/GSC all-sky survey and pulse period data
from the Fermi/GBM pulsar project.
X-ray binaries, consisting of a highly magnetized neutron star
and a stellar companion, originate X-ray emission according to 
the energy of the accretion matter onto the neutron star.  
The accretion matter also transfers the angular momentum at the Alfven radius, 
and then spin up the neutron star.  Therefore, the
X-ray luminosity and the spin-up rate are supposed to be well
correlated.  
%
%
%
We analyzed the luminosity and period-change relation using 
the data taken by continuous monitoring of MAXI/GSC and Fermi/GBM 
for Be/X-ray binaries, GX 304$-$1, A 0535$+$26, GRO J1008$-$57, KS
1947$+$300, and 2S 1417$-$624, which 
occurred large outbursts in the last four years.
We discuss the results comparing the obtained observed relation with that of 
the theoretical model by Ghosh \& Lamb (1979).

\end{abstract}

\maketitle

\thispagestyle{fancy}


\section{Introduction}

X-ray binary pulsars (XBPs) are systems consisting of magnetized
neutron stars and mass-donating stellar companions. 
%
%
Since the neutron stars are strongly magnetized, the matter flows from
the companion are dominated by the magnetic pressure inside the
Alfven radius, and then funneled onto the magnetic poles along
the magnetic field lines.  
The accretion matter also transfers its angular momentum at the
Alfven radius.
Therefore, the pulsar spin-up rate and the
mass accretion rate, i.e. the X-ray luminosity, are thought to be
closely correlated (e.g. Ghosh \& Lamb 1979, hereafter GL79 \cite{Ghosh_Lamb1979}).
The issue is relevant to the fundamental parameters of 
the neutron stars such as mass, radius, and magnetic field,
as well as the XBP evolution scenarios.


Be XBPs, which include Be stars extending circumstellar disk around
the equator, are one of the major XBP subgroups \cite{Reig2011}.  They
often exhibit large outbursts lasting for about a few weeks to a few
months mostly at around the orbital phase of the neutron-star
periastron passage.  During these outbursts, simultaneous spin-up
episodes are often observed (e.g. \cite{1997ApJS..113..367B}).  This
is naturally explained by an increase in the accretion rate 
induced by the interaction with Be-star disk, and the
associated transfer of the angular momentum to the neutron star via
disc-magnetosphare coupling.
These events give us an opportunity to study the 
relation between the luminosity and the spin-up rate
quantitatively.

In this paper, we present the study on the relation using the
long-term light curve obtained by the MAXI/GSC all-sky survey and the
period change obtained from the archived results of Fermi/GBM pulsar
project.
These data, taken by the continuous monitor for over four years, enable
us to investigate their time variations over the entire outburst
activities in Be XBPs.
We describe the observation in $\S$ \ref{sec:obs_data}, the analysis
procedure in $\S$ \ref{sec:analysis}, and then discuss about the
obtained results in $\S$ \ref{sec:discussion}.

\section{Observation Data}
\label{sec:obs_data}



Since the MAXI (Monitor of All-sky X-ray Image;
\cite{Matsuoka_pasj2009}) experiment onboard the International
Space Station started in 2009 August, 
the GSC (Gas Slit Camera; \cite{Mihara_pasj2011}),
one of the two MAXI detectors, has been scanning 
almost the whole sky every 92-minute orbital cycle 
in the 2--30 keV band.
To obtain the long-term luminosity variation of 
Be XBPs covering the outbursts as well as the intermission/quiescence, 
we use archived GSC light-curve data 
in 2--20 keV band,
which are processed with a standard procedure \citep{sugizaki_pasj2011}
by the MAXI team 
and archived at MAXI web site\cite{maxi_home-ref}.


The GBM (Gamma-ray Burst Monitor; \cite{Meegan2009}) onboard the Fermi
Gamma-Ray Space Telescope, is an all-sky instrument sensitive to
X-rays and gamma-rays with energies between 8 keV and 40 MeV.  The
Fermi GBM pulsar
project \cite{2009arXiv0912.3847F, 2010ApJ...708.1500C}
provides results of timing analysis of a number of positively detected
X-ray pulsars, including their pulsation periods and pulsed fluxes
via the web site \cite{fgbm_pulsar-ref}
since the in-orbit operation started in 2008 July.
We utilized the archived pulse period data of Be XBPs.

\begin{figure*}
\begin{center}
\begin{minipage}[t]{78mm}
\includegraphics[width=78mm]{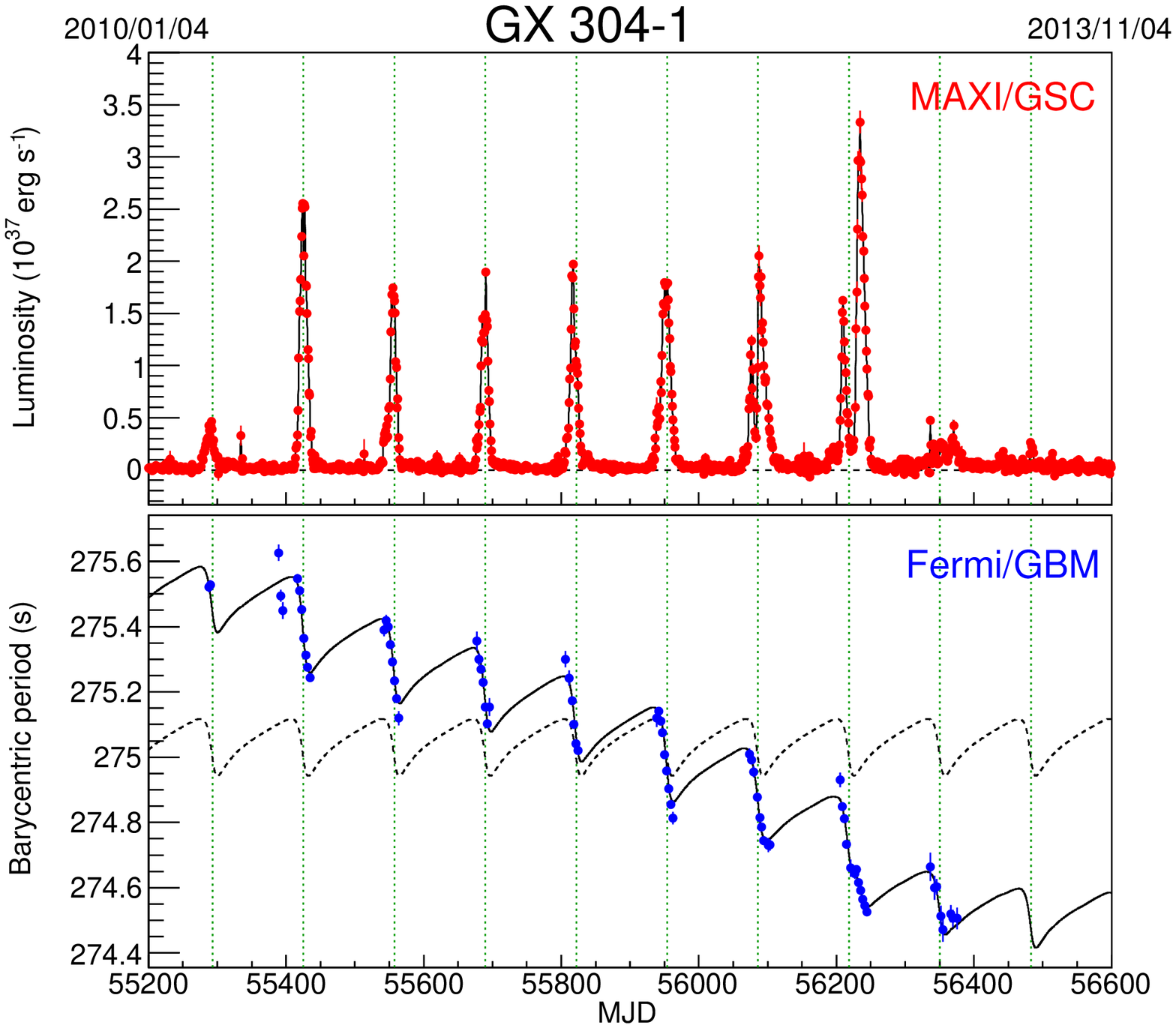}
\vspace{3mm}
\includegraphics[width=78mm]{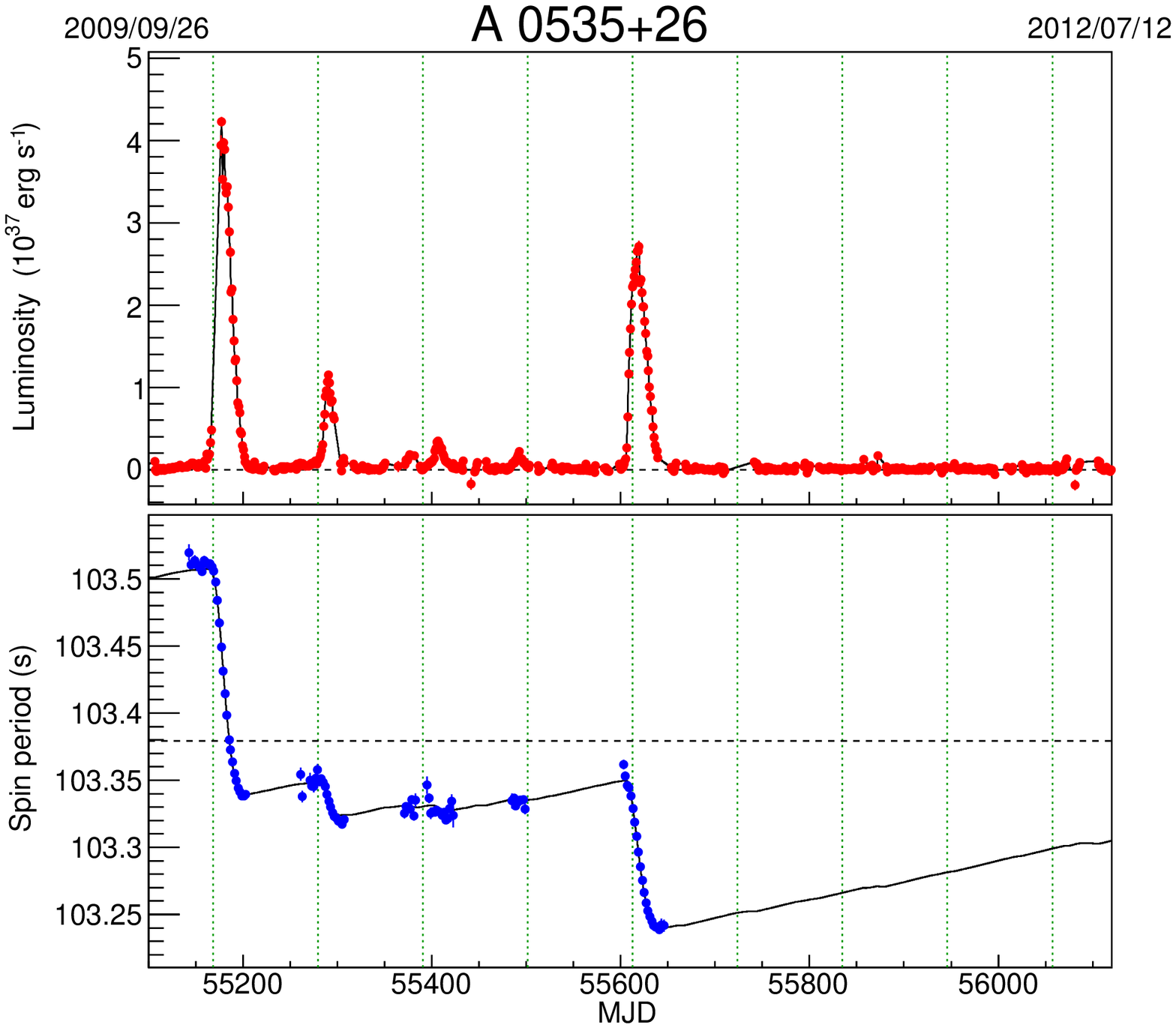}
\vspace{3mm}
\includegraphics[width=78mm]{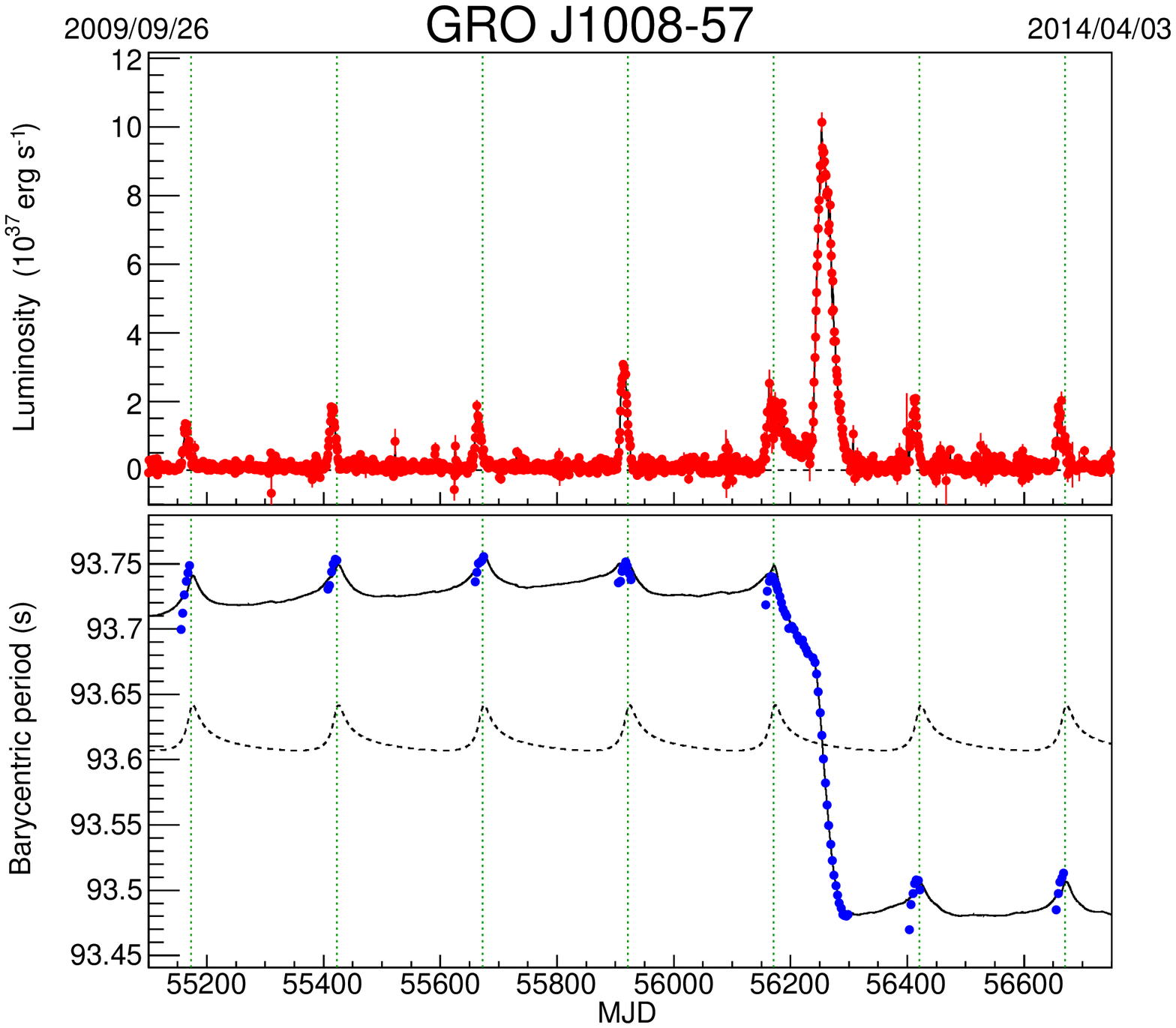}
\end{minipage}
\begin{minipage}[t]{78mm}
\includegraphics[width=78mm]{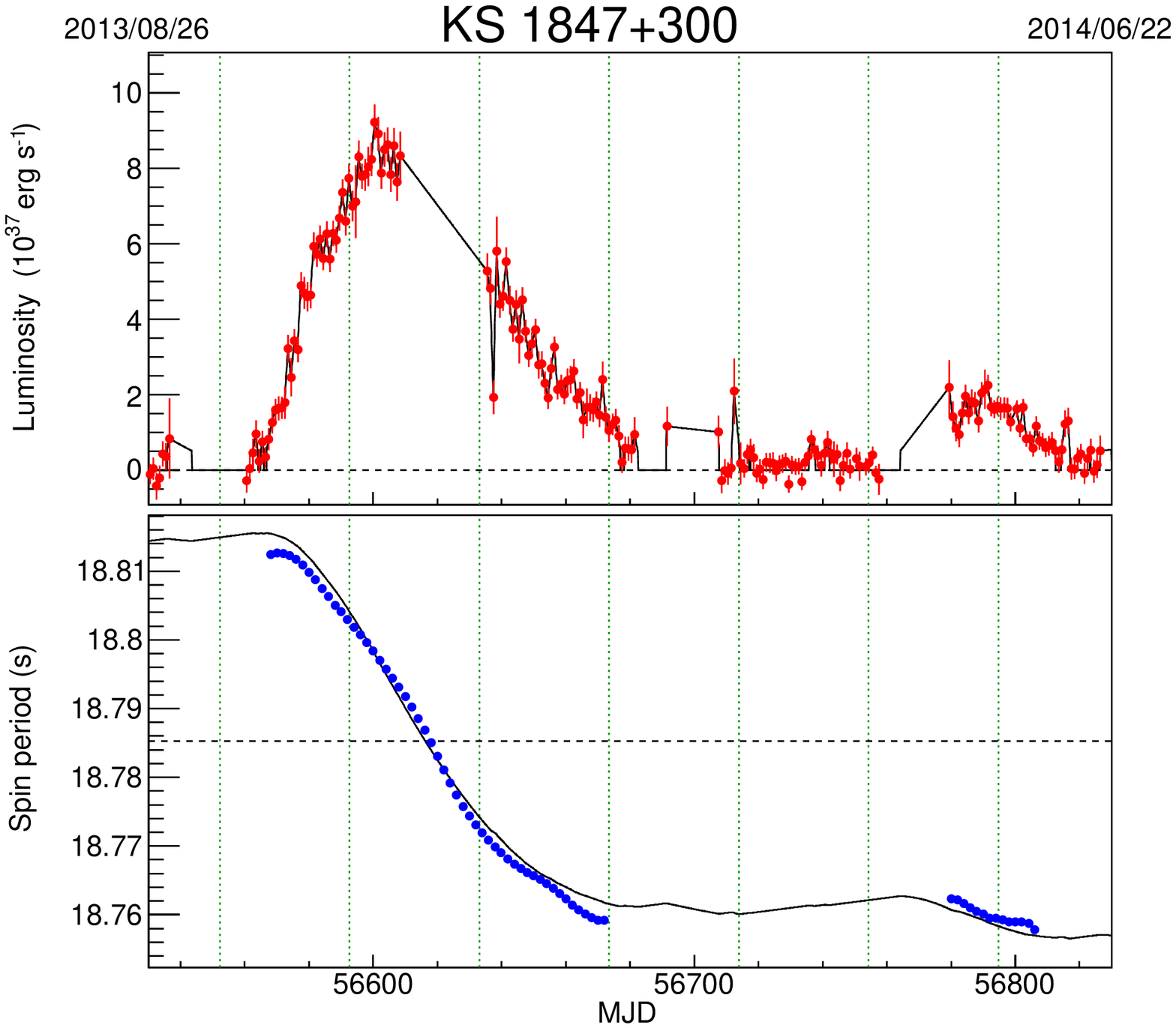}
\vspace{3mm}
\includegraphics[width=78mm]{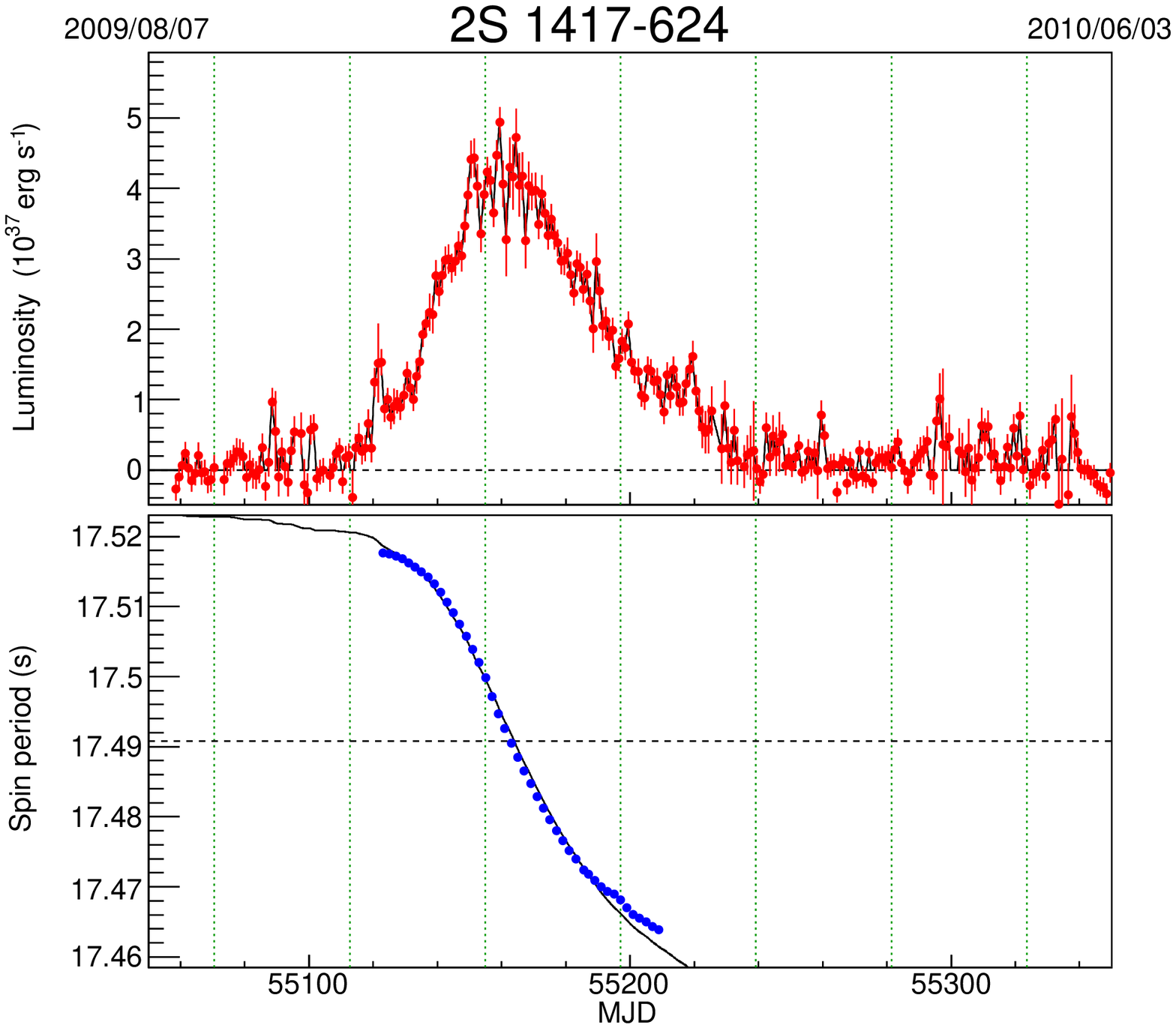}
\end{minipage}
\caption{
For each target of
GX 304$-$1,     
A 0535$+$26, 
GRO J1008$-$57, 
KS 1947$+$300, and   
2S 1417$-$624, 
time variation of luminosity estimated from
MAXI/GSC 2--20 keV light curve data in 1-d time bin (top) and 
and that of pulse period during the outbursts 
obtained from Fermi/GBM pulsar data (bottom)
are plotted.
In the top panels, 
solid lines represent the luminosity data $L_{37}(t)$ 
used for the period-change model fit.
In the bottom panel,
solid and dash lines represent the best-fit period model
and the inclusive orbital Doppler effects which 
have been corrected 
in A 0535$+$26, KS 1947$+$300, and  2S 1417$-$624.
}
\label{fig:lx_period_fit}
\end{center}
\end{figure*}

We selected five Be XBPs, GX 304$-$1, A 0535$+$26, GRO J1008$-$57, KS
1947$+$300 and 2S 1417$-$624
from targets listed in the MAXI/GSC and the Fermi/GBM archive
for this study,
because 
they exhibited large outburst activities in the last four years and
their surface magnetic fields are
well determined by the cycloton resonance feature in the X-ray spectrum
(execpt for 2S 1417$-$624).
Table \ref{tab:beparam} summarized
characteristic parameters of these Be XBPs
and figure \ref{fig:lx_period_fit} shows the time variation of the
bolometric luminosity calculated from MAXI/GSC 2--20 keV light curve
data 
and that of the pulse period obtained from the
Fermi/GBM pulsar data during outbursts for each source.


\section{Analysis}
\label{sec:analysis}

Observed pulse-period variations of XBPs include two distinct effects,
the intrinsic pulsar spin-period change and the orbital Doppler
effect.
In Be XBPs,
both of them are supposed to correlate with the orbital phase.
Therefore, it is not straightforward to resolve each component
from the observed data.
Although
the pulse period data of XBPs in the Fermi/GBM archive are
corrected for the orbital Doppler effect if their orbital elements are
determined, the orbital elements have not been known in
all of the Be XBPs with our interests.
Hence, we construct a semi-empirical model implementing both these effects
and then fit it to the data, in an attempt to
simultaneously determine the intrinsic pulse
period change and the orbital elements.

\subsection{Modeling of period change in XBPs}

We here employ the simple theoretical model of the pulsar spin-up
by the mass accretion via disk, proposed by
GL79 \cite{Ghosh_Lamb1979}.
The model has been examined with X-ray data,
and its validity and limits are well studied
(e.g. \cite{1996AA...312..872R, 1997ApJS..113..367B}).
In this model, 
the pulsar spin-up rate $-\dot{P}_{\rm spin}$ (s yr$^{-1}$) is given by
\begin{equation}
-\dot{P}_{\rm spin}  =  5.0 \times 10^{-5} \mu_{30}^{2/7} n(\omega_{\rm s})  
S_{1}(M) P_{\rm spin}^2 L_{37}^{6/7} \label{equ:pdot}
\end{equation}
\[
S_{1}(M) = R_{\rm 6}^{6/7} (M/M_{\odot})^{-3/7}I_{45}^{-1} 
\]
%
%
where 
$\mu_{30}$, $R_{\rm NS6}$, $M_{\rm NS\odot}$, $I_{45}$, $P_{\rm spin}$, $L_{\rm 37}$ 
are the magnetic dipole moment of the neutron star
in units of $10^{30}$ G cm$^{3}$, 
radius in $10^6$ cm, 
mass in $M_\odot$, 
moment of inertia in $10^{45}$ g cm$^2$,
spin perid in s, 
luminosity in $10^{37}$ erg s$^{-1}$,
$n(\omega_{\rm s})$ is a dimensionless torque that
depends on the fastness parameter $\omega_{\rm s}$
and approximately constant at $\sim 1.4$ in slow rotating pulsars
satisfying $(P_{\rm spin}L_{\rm 37}^{3/7}) \gg 1$.

The equation \ref{equ:pdot} implies that the spin-up rate $-\dot{P}_{\rm spin}$
follows the luminosity $L$ 
as $-\dot{P}_{\rm spin}\propto L^{6/7}$.  
%
%
The power-law index $\gamma$ in a model of $-\dot{P}_{\rm spin}\propto L^{\gamma}$  
obtained from the fit to the observed data sometimes disagreed
with the theoretical value of $6/7$ and favor the rather
higher value of $\sim 1.2$ \citep{1996AA...312..872R, 1997ApJS..113..367B}.
%
Besides this, the comparison of absolute spin-up rate with
equation \ref{equ:pdot} has been hampered by a large
uncertainty in the bolometric luminosity correction,
which is in turn due to beaming effects 
(e.g. \cite{1997ApJS..113..367B,  2002ApJ...570..287W}).
We hence employ the spin-up model expressed by
\begin{equation}
-\dot{P}_{\rm spin} = \alpha L_{37}^\gamma
\end{equation}
in which the power-law index $\gamma=6/7$ and
a correlation factor, 
$\alpha = 1.7 \times 10^{-7} \mu_{30}^{2/7} P_{\rm spin}^2$ s d$^{-1}$ ($=\alpha_0$)
reduced from the equation \ref{equ:pdot}
and typical neutron-star parameters of $R_6=1$, $M=1.4M_\odot$, $I_{45}=1$, 
are treated as free parameters.


XBPs are also known to spin down during the quiescence due to
the propeller effects.  
The rate is much smaller than the spin-up
during the outburst bright phases, but may not be negligible.
We accounted its effect with a constant spin-down parameter, $\beta$, 
added to $\dot{P}_{\rm spin}$ as an offset.

By combining the spin-up and spin-down models above, 
the intrinsic pulsar-spin period $P_{\rm spin}(t)$ is expressed by
\begin{eqnarray}
P_{\rm spin}(t) &=& P_0 + \int_{\tau_0}^t \dot{P}_{\rm spin}(\tau) d\tau \nonumber \\
&=& P_0 + \int_{\tau_0}^t  \left\{ -\alpha  L_{37}^{\gamma}(\tau) +\beta\right\}d\tau 
\label{equ:pspinmod}
\end{eqnarray}
%
%
where we set the time basis $\tau_0$ at the first periastron passage
in the period under analysis and define the pulsation period at the time
$\tau_0$ as $P_0=P_{\rm spin}(\tau_0)$ .
The model equation \ref{equ:pspinmod} 
includes four free parameters, $P_0$, $\alpha$, $\beta$, $\gamma$ and 
requires the luminosity data $L_{37}(t)$ 
as a function of time.
We calculated the luminosity from data of
the MAXI/GSC 2--20 keV light curve in 1-d time bin assuming the source
distance, the typical energy spectrum of a cutoff power
law from the past results, and the source emission to be isotopic.

The period modulation due to the binary orbital motion is calculated by using
the binary elements, which consists of orbital period $P_{\rm B}$,
eccentricity $e$, projected semi-major axis $a_{\rm x}\sin i$, epoch
$\tau_0$ and argument $\omega_0$ of the periastron.
The pulsar orbital velocity $v_{\rm l}(t)$ along the line of sight is 
\begin{equation}
v_{\rm l}(t)
= \frac{2\pi a_{\rm x} \sin i}{ P_{\rm B} \sqrt{1-e^2}} \left\{\cos\left(\nu(t)+\omega_0\right)+e\cos\omega_0\right\}
\label{equ:vl}
\end{equation}
where $\nu(t)$ is a parameter called 'true anomaly' describing the
motion on the elliptical orbit and calculated from the Kepler's
equation.
The observed pulse period, $P_{\rm obs}(t)$, is then expressed by
\begin{equation}
P_{\rm obs}(t) \simeq P_{\rm spin}(t) \left(1+\frac{v_{\rm l}(t)}{c}\right).
\label{equ:pobsmod}
\end{equation}

\begin{table*}[bht]
\begin{center}
\caption{
Characteristic parameters of selected Be X-ray binary pulsars and the best-fit parameters ($\alpha$, $\beta$)
used in the period change model.
}
\label{tab:beparam}
\begin{tabular}{lcccccc@{\hspace{8mm}}cc@{\hspace{8mm}}c}
\hline
\hline
Target name    & $P_{\rm pulse}$ & $P_{\rm orbit}$ & $a_{\rm X} \sin i$ & $e$ & $B$ & $D$ & $\alpha/\alpha_0$ &  $\beta$ & Ref. \\
               & ( s ) & ( d ) & ( lt-s ) &  & ( $10^{12}$ G ) & ( kpc ) & & ( $10^{-9}$ s s$^{-1}$ ) & \\
\hline
GX 304$-$1     & 275 &  132.19 & 500 & 0.5~~  & 4.7  & 2.0  & 0.28  & 2.0~~~~   & \cite{2011PASJ...63S.751Y,2014efxu.conf..226S} \\
A 0535$+$26    & 103 &  111.10 & 267 & 0.47~  & 4.3  & 2.4  & 1.3~  & 3.6~~~~   & \cite{2013ApJ...764L..23C,2013AA...552A..81M}  \\
GRO J1008$-$57 & 93  &  249.48 & 530 & 0.68~  & 6.6  & 5.8  & 0.49  & 2.5~~~~   & \cite{2013AA...555A..95K,2014PASJ...66...59Y}  \\
KS 1947$+$300  & 18  &  ~40.42 & 137 & 0.034  & 1.1  & 10~~ & 3.2~  & 0.69~~~   & \cite{2004ApJ...613.1164G,2014ApJ...784L..40F}  \\
2S 1417$-$624  & 17  &  ~42.18 & 188 & 0.44~  & --$^*$   & 11~~ & (6.8)$^*$  & 0.0:fix   & \cite{1996AAS..120C.209F,2004MNRAS.349..173I}  \\
\hline
\end{tabular}
\end{center}
$^*$: The surface magnetic field $B$ has not been measured. It is assumed to be $2\times 10^{12}$ G.
\end{table*}

\subsection{Period-change model fit}
\label{sec:results}

We applied the spin-period-change model, $P_{\rm spin}(t)$ in equation
\ref{equ:pspinmod}, to the Fermi/GBM archived period data for A
0535$+$26, KS 1947$+$300, and 2S 1417$-$624, in which the binary
orbital effects were corrected with the known orbital elements.
About GRO J1008$-$57, 
the orbital effects are not corrected in the archived data,
but the orbital elements have been estimated by
\cite{2013AA...555A..95K}. We thus fit the data to the period model,
$P_{\rm obs}(t)$ in equation \ref{equ:pobsmod}, which includes the orbital effect,
employing the orbital elements given in \cite{2013AA...555A..95K}.
About GX 304$-$1, its orbital elements have not been
measured. We fit the period data with
the model $P_{\rm obs}(t)$ in which the orbital elements are floated.

As results of many model-fit attempts,
we found that the model is able to reproduce the data 
approximately with $\gamma\sim 1$ in all of the five targets.  
We thus fix the parameter $\gamma$ at $6/7$, predicted by GL79 \cite{Ghosh_Lamb1979},
in order to concentrate on the correlation factor $\alpha$, hereafter.
In figure \ref{fig:lx_period_fit} bottom panels, 
the obtained best-fit models with $\gamma=6/7$ are superposed on the period data.
The best-fit parameters are shown in table \ref{tab:beparam},
where the values of $\alpha$ are given by the ratio to
that ($=\alpha_0$) predicted by GL79 \cite{Ghosh_Lamb1979}.


\section{Discussion}
\label{sec:discussion}

We fitted pulse period variation of five Be XBPs observed with Fermi/GBM
to the model implementing the spin-up due to 
the mass accretion via disk,
expressed by $\dot{P}_{\rm spin} = \alpha L^{6/7}$
based on GL79 \cite{Ghosh_Lamb1979}
and the luminosity estimated from the MAXI/GSC light curve.
The results show that the model successfully reproduce the data
in all of the five samples.
The obtained best-fit parameters imply that the correlation
factor $\alpha$ from the luminosity $L^{6/7}$ to the spin-up rate
$\dot{P}$ largely agree with $\alpha_0$ predicted by GL79
\cite{Ghosh_Lamb1979}. 
%
The dispersion of the ratio, $\alpha/\alpha_0 \sim0.3$ to $3$,
is naturally expected from the uncertainty in the
bolometric luminosity correction due to the beaming effect.

However, the values of $\alpha/\alpha_0$ seems to 
have some tendency against the pulse period, the orbital period, 
and the eccentricity, which are suggested to have a relation
with Be-XBP subgroups \cite{2011Natur.479..372K}.
This will become clearer with increasing data
in the near future.

%
%

\bigskip 
\begin{acknowledgments}

The authors thank all of the MAXI team members for 
enabling the data analysis for the science study
as well as the Fermi/GBM pulsar project for providing 
the useful analysis results to the public.

\end{acknowledgments}

\bigskip 


\end{document}